\documentclass[12pt,thmsa]{article}
\usepackage{amsfonts}
\usepackage{color}

\begin{document}

\author{Karl-Georg Schlesinger \\
J{\"u}lich Supercomputing Centre (JSC)\\
Forschungszentrum J{\"u}lich GmbH\\
D-52428 J{\"u}lich, Germany\\
e-mail: kggschles@gmail.com}
\title{Notes on the firewall paradox, complexity, and quantum theory}
\date{27.11.2014}
\maketitle

\begin{abstract}
We investigate what it means to apply the solution, proposed to the firewall paradox of \cite{AMPS} by \cite{HH}, to the famous quantum paradoxes of Schr{\"o}dinger's Cat and Wigner's Friend if ones views these as posing a thermodynamic decoding problem (as does Hawking radiation in the firewall paradox). The implications might point to a relevance of the firewall paradox for the axiomatic and set theoretic foundations underlying mathematics. We reconsider in this context the results of \cite{Ben1976a} and \cite{Ben1976b} on the foundational challenges posed by the randomness postulate of quantum theory. A central point in our discussion is that one can mathematically not naturally distinguish between computational complexity (as central in \cite{HH} and \cite{Sus}) and proof theoretic complexity (since they represent the same concept on a Turing machine), with the latter being related to a \textit{finite} bound on Kolmogorov entropy (due to Chaitin incompleteness).

\end{abstract}

\bigskip

\section{Introduction}

The firewall paradox, originally posed in \cite{AMPS}, shows -- under very general assumptions -- the incompatibilty of the unitarity requirement of quantum mechanics with the equivalence principle of general relativity even for weak gravitational fields. The argument fundamentally rests on the idea that one could use the Hawking radiation of a black hole to read information on the interior from it, before falling in. It is the revolutinary achievement of \cite{AMPS} to show this to be actually possible in an intricate thought experiment. In \cite{HH} a possible solution to the paradox was proposed by showing that the decoding of the Hawking radiation is a process of high computaional complexity (in the sense of algorithmic information theory), in consequence taking too long a time scale to be performed before the evaporation of the black hole. The solution proposed in \cite{HH} -- and further pursued in \cite{Sus} -- can be seen as trying to formulate the most conservative way out. For this reason -- since the alternative would be to accept a brakedown of physics, as known so far, even for weak gravitational fields -- it seems worthwile to stay with this approach as long as possible. Since the proposal of \cite{HH} fundamentally rests on the complexity of decoding the Hawking radiation, one can pose the question if paradoxes of this kind are really limited to the setting of gravitation or if they point at a deeper conflict of quantum mechanics with complex thermodynaic systems.\\
The famous quantum paradoxes of Schr{\"o}dinger's Cat and Wigner's Friend can also be seen as posing a decoding problem of a thermodynamic type (as in the case of Hawking radiation). We investigate what it means to apply the complexity theory based solution, as proposed to the firewall paradox of \cite{AMPS} by \cite{HH}, also to these situations. The implications might point to a relevance of the firewall paradox for the very foundations of mathematics, in the form of axiomatic systems and set theory. To deepen the perspective, in subsequent sections we reconsider these implications from the viewpoint of consequences which Chaitin incompleteness (see \cite{CC} and \cite{Cha}) has in quantum theory, as well, as from the old results of \cite{Ben1976a} and \cite{Ben1976b} on implications of the randomenss postulate of quamtum mechanics for the set theoretic foundations of mathematics. In an appendix, we reconsider the implications from a more physics based perspective.

\section{Schr{\"o}dinger's Cat}

The famous paradox of Schr{\"o}dinger's Cat is often viewed from a more or less philisophical perspective as stressing the strangeness of the quantum mechanical superposition principle, by bringing it to play on the notions of an alive or dead cat. Strange as the superposed state

\begin{equation}
|{\Psi}> = \frac{1}{\sqrt{2}}(|alive>+|dead>)
\end{equation}

is, it has the even stranger consequence that one can use it to bring dead cats back to live. By the fundamental rules of quantum mechanics, there exists a self-adjoint operator $A$ having $|{\Psi}>$ has an eigenvector. Starting from the state $|dead>$, describing a dead cat, you can choose to measure the obeservable $A$. The outcome is with probability $\frac{1}{2}$ the state $|\Psi>$ and with probability $\frac{1}{2}$ the state

\begin{equation}
|{\Phi}> = \frac{1}{\sqrt{2}}(|alive>-|dead>)
\end{equation}

with the opposite superposition. But starting from the eigenstate $|{\Psi}>$ (and likewise for $|{\Phi}>$), you can choose to measure (in this case by simply having a look at it) if the cat is alive or dead. Once again, you will get the result with probability $\frac{1}{2}$ in each of the eigenvectors $|alive>$ or $|dead>$. In consequence, by performing these two measurements in succession, you have a nonvanishing probability ($\frac{1}{2}$, using both transitional states $|{\Psi}>$ and $|{\Phi}>$) of starting from the state $|dead>$ and ending up with the state $|alive>$, bringing back the dead cat to life in this way. But the important point is that this thought experiment is not just strange (which could equally well just result from our prejudice with concepts as alive and dead, as resulting from experience with classical instead of quantum objects, as it could point to problems in our understanding of quantum mechanics) but has real physical content. The transition from life to death is -- in prosaic terms of physics -- definitely a transition involving an increase in entropy (all the information contained in neuronal activity and much more is lost). But the state $|alive>$ having lower entropy as the state $|dead>$ means, by performing the two successive measurements, as described above, you can \textit{at any time} (since the axioms of quantum mechanics presuppose that all self-adjoint operators are available for measurement at any time) purposfully (nearly, i.e. with the -- in statistical mechanics terms rather high -- probability $\frac{1}{2}$) decrease the entropy of the physical system given by the cat.  \\
You might argue against this conclusion by stating that one can not use a desription of the system in terms of pure quantum mechanical states and as a thermodynamic system at the same time since the latter relies e.g. on some form of coarse graining and would on the quantum side have to correspond to a density matrix state. In other words, the reduction in entropy would just result from removing the coarse graining on passing to a pure state description, i.e. the paradoxical conclusion would be just a kind of quantum version of the Laplacian demon. I do not think this is a valid counter argument. What Schr{\"o}dinger does in his thought experiment is starting from the premise that we can \textit{in principle} describe even the macroscopic system of the cat in terms of a pure quantum mechanical system. Hence, the operator $A$ exists (which is the crucial step to apply the two successive measurements as described above). The situation is rather different from classical mechanics were passing to a many body description (and ultimately statistical ensembles) does in no way interfer with the basic Hamiltonian structure (contact transformations, etc.) of classical mechanics. In quantum mechanics the situation is different: Once we have accepted a quantum mechanical description even for the whole macroscopic system, the superpositions and the operator $A$ are undeniably there by the very axioms of quantum mechanics. On the other hand, the macroscopic system of the cat has the well known classical description in terms of alive or dead which definitely has a thermodynamic content in terms of emtropy. I strongly believe that Schr{\"o}dinger is pointing to exactly this point with his paradox. \\
To phrase the situation differently, imagine you decide to observe the situation of the cat together with a friend. What Schr{\"o}dinger points to is that your friend could decide to describe the cat as a completely quantum mechanical system, including the construction of the operator $A$. At the same time, you could decide to consider the cat the way you have ever done. This latter description includes the increase in entropy in the transition from alive to dead cat. The paradox arises once you grant your friend the possibility to describe -- \textit{in principle} -- the complete macroscopic cat as a quantum mechanical system. You can not go on to later forbid your friend a single state description by pointing to coarse graining as the origin of your own description because that would mean that systems which are described in essential features by properties resulting from coarse graining do -- even \textit{in principle} -- not allow for a single state description. On the other hand, if you grant your friend the single state description -- as Schr{\"o}dinger does -- this does in no way touch the fact that you will observe the cat as usual, including the increase in entropy. \\  
Now, let us suppose that a similar solution as proposed in \cite{HH} for the firewall paradox could also be pursued for the case of Schr{\"o}dinger's Cat. We should strongly stress that we use this on the level of an \textit{assumption}. Even if it would be applicable, it could still be a long way to rigorously prove such an argument in the framework of complexity theory for a puerly quantum mechanical system. The assumption is that -- as in the case of decoding the Hawking radiation -- it would be a process of high computational complexity -- consequently taking a very long time -- to actually realize the operator $A$. Observe that constructing the operator $A$ \textit{is} similar to decoding thermal radiation since it involves rereading the information of the alive cat from the dead one (as the two stage experiment bringing it back to life shows since only the measurement of $A$ is involved, the final measurement basically consits of just having a look at the cat). To make the analogy even more compelling we could sligthtly modify the thought experiment by not putting the cat into a box with a poison but observing it by astronomical techniques while the cat is placed on another planet. There the radiation quantum would not simply trigger the offset of a poison but the explosion of an atomic bomb. This does not change the argument of the paradox at all but it shows that the state of the dead cat would largely get the character of thermal radiation.\\
Let us further assume that -- as in the solution to the firewall paradox proposed in \cite{HH} -- it takes a time on the order of the recurrence time of the system under consideration -- so, in this case the cat -- to construct the operator $A$. This looks immediately promising to resolve the paradox since it is precisley the recurrence time under which the sytem is thermodynamically allowed to reduce its entropy because it simply returns to its orginal microscopic state by chance. \\
But we have to be careful at this point because one has to decide between the classical Poincare recurrence time versus the quantum recurrence time of the system. It is the quantum recurrence time which is the relevant one for the system to return by chance to its original state and, hence, for the reduction of entropy. Thermodynamics is not conistent with an underlying classical theory as we know especially well for the case of radiation from Planck's historic discovery. It is the quantum recurrence time under which Hawking radiation would by chance reensemble into a classical black hole (see \cite{HH}. The recent discovery that mitochondria probably involve EPR quantum correlations, in order to explain their high energy efficiency, further stresses that we can not expect to read off the informtion of the alive cat without invoking information from quantum correlated states, further stressing the quantum recurrence time as the relevant one. \\
What about the complexity of $A$, i.e. the time it would take to construct it in a quantum computation. If we keep to the analogy to \cite{HH}, this reduces from the -- exponentially larger -- quantum recurrence time to classical Poincare recurrence time, provided we can use the quantum circuit model for the quantum computation which essentailly boils down to the requirement that we can radiate away random information into a thermodynamic heat reservoir. Can we grant this for the case of Schr{\"o}dinger's Cat? The answer is yes because the needed heat reservoir is outside the system of the cat. The situation is very much as in another famous thought experiment, known as Wigner's Friend. In this case, a person (Wigner's Friend) is the observer for the box with the cat. He is the one who performs the quantum computation to construct the operator $A$ and he is the one who has to have the means of radiating away information into a thermodynamic resrvoir. As long as we grant Wigner's Friend to be a system described by classical phyiscs, this does absolutely cause no problem. \\
But then we can not resolve the paradox of Schr{\"o}dinger's Cat. We can pass from a more than exponential contradiction (between availability of $A$ at any time in quantum mechanics and the quantum recurrence time for the sytem to reduce its entropy by chance) to an exponetial one (between classical Poincare and quantum recurrence time) but still a contradiction in terms of a huge discrepancy of time scales remains. This would only be removed if we would not grant Wigner's Friend access to a thermodynamic reservoir in which case the time he would need to complete the quantum computation would also scale up to the quantum recurrence time. But this would mean, we would treat Wigner's Friend himself as a quantum mechanical sytem, involving superpositions, etc..\\

\bigskip

Observe that the difference between the classical and the quantum description of Wigner's Friend -- resulting in classical Poincare versus quantum recurrence as the time sacle for the quantum computation -- is esentially a difference in information available to this system at any time, as shown by the crucial relevance of the possibility to radiate information away. We will take a more detailed look on Wigner's Friend in the subsequent section.

\section{Wigner's Friend}

Should Wigner's Friend itself be described as a quantum mechanical system with superposed states? Here, the definite answer seems to be that this is not possible. It would mean that scientific results, the analysis of experiments, publishing of papers about results, and all this, would live in a superposed state until \textit{me} personally coming into the lecture hall of some conference venue and reducing everything to a classical description. We can not pass to such a description because it would undermine one of the basic assumptions behind science, putting all participants in the scientific enterprise on the same footing and viewing outcomes of experiments as classical \textit{facts} which we can discuss as such in papers. Without a classical thermodynamic world (involving the second law) there would be no notion of \textit{facts}. Bohr always stressed this necessity of a classical description of the final observer because we have to be able to communicate results to and discuss them with colleagues. \\
It is the indeterminacy of quantum superpositions which leads us to requiring a classical description for Wigner's Friend. Though intuitively plausible, it is not clear if this inderterminacy does agree with the difference in accessible information for a classical versus a quantum computing system, as exemplified in the difference between classical Poincare and quantum recurrence times. Even if one could show mutual implication for the case of conventional nonrelativistic quantum mechanics, we do already know that there is no agreement on the level of general principles from groundbreaking results of \cite{AW}. It is shown there that in the strange world of closed timelike curvers (CTCs, for short) -- involving some form of a consitency condition, of course, since nobody should be able to kill his grandfather -- the notions of classical and quantum computability agree, i.e. there is no difference in accessible information any more. More concretely, for both -- classical and quantum Turing machines -- the very large complexity class PSPACE (which includes and is believed to properly include the class NP) becomes accessible in polynomial time. \\
In consequence, in such a world the requirement that Wigner's Friend obeys to a classical description without superpositions -- in accordance with the requirement of Bohr that results of experiments have to be communicable in the community of scientists -- would not lead to a contradiction in information accessible to classical versus quantun systems and, therefore, not a contradiction between computing times. The issue of classical Ponicare versus quantum recurrence time taken for the computation would be resolved. 

\bigskip

A person using a mathematical description in accordance with the classical and quantum Turing machines of \cite{AW} would see PSPACE as a class of polynomial computing time, i.e. in such a mathematical description we would have 

\begin{equation}
P=PSPACE
\end{equation}      

Remember, here, that a different Turing machine just means a different formulation of mathematics. In consequence, a different type of Turing machine requires a different -- compatible with this notion of Turing machine -- formulation of set theory (as the foundational underpinning of mathematics), an observation already made by John v. Neumann. In this case the solution to the paradoxes would not be the conservative one, proposed for the firewall paradox by \cite{HH}, but the conclusion would be that we are hindered in our understanding of quantum theory and black holes/holography by the foundations our present formulation of mathematics is build on (still this can be seen as a conservative proposal compared to a brakedown of physics, as we presently know it, even for weak gravitational fields). Let us explain this in more detail.\\
A mathematical description, in accordance with the classical and quantum Turing machies of \cite{AW}, requiring $P=PSPACE$ leaves us with three possibilities: The first is that this holds anyway in our present foundations of mathematics (given by the set theory ZFC or some mild modification thereof). This is still possible but not the expectation held by most researchers in the field. The second possibility is that the statement is actually independent -- as an example of G{\"o}del incompleteness -- from the axiom system ZFC. In this case, one could include (3) as an additional axiom, i.e. one would be lead to an extension of set theory beyond the power of ZFC. The third case (the one most experts in the field would expect) means that (3) can actually be disproved from the axioms of ZFC. This would mean that we could not get a new foundational set theory by simply passing to an extension of ZFC, in order to incorporate (3), but that we would have to use a different set theory altogether. The fact that this theory would have axioms in contradiction with ZFC does not make it an absurdity. Of course, we would have to require that large parts of our present day mathematics could be formulated successfully also in this alternate system. From the side of physics, one would, of course, have to require that e.g. the mathematical machinery for general relativity, quantum mechanics, and quantum field theory would still be available in this system. Though these are strong requirements, there is no reason to assume a priori that they are impossible to meet. \\
Let us take a short look at some examples from the realm of axiomatic set theory which can serve to explain this view in more detail:

\bigskip

\begin{itemize}

\item The modern formulation of the concepts of calculus and the formulation of Newton in terms of infinitesimals are rather different and for a long time the infinitesimals have been just seen as an intuitive and calculationally attractive approach, lacking logical consistency. As, is well known, this is not true and the approach based on infinitesimals can also be given a logically sound basis (this approach came to be known under the name of nonstandard analysis). It is possible to completely avoid the model theoretic complications, involved in the first approaches to this subject, and base nonstandard analysis completely in an alternate axiomatic set theory (IST, for internal set theory, see \cite{Nel}), just as usual analysis is based in ZFC. So, mathematics and physics can be done as usal -- though with a different flavour -- in IST but IST and ZFC differ strongly in the deeper transfinite realm (though IST is an extension of ZFC, i.e. they are not in contradiction to each other). This example shows that large parts of mathematics and the applications to mathematical phyiscs do not determine transfinite structures to an extent which would make the foundational choice of ZFC unique (There is a later formulation of nonstandard analysis in terms of an alternate axiomatic set theory called BST -- for bounded set theory -- which shows BST to be just a kind of language trick played on ZFC, in this way removing the differences in transfinite structures between IST and ZFC from the foundations of nonstandard analysis, see \cite{KR1995a}and \cite{KR1995b}. Though this was an important step in the foundations of nonstandard analysis, it does in no way affect the use of the older IST approach as an example in our context, here.).  

\item Another example is offered by alternate set theories placed in the framework of topos theory (see \cite{Gol} for an introductory overview). These using the intuitionistic propositional calculus of Heyting algebras -- instead of the usual Boolean algebra based propositional calculus -- shows another possible feature of alternate set theories: In search for alternate foundations one has to take into account all three blocks of axioms (the axioms of the set theory proper, the basic logical axioms, and the inference rules) together and can not a priori limit the change to one of the blocks. 

\item Our final example shows that even if the requirement $P=PSPACE$ would turn out to establish a contradiction with axioms of ZFC used in the foundation of e.g. general relativity or quantum mechanics, this is not necessarily the end of the road. Kunen inconsitency is generally viewed as an upper limit to the pyramid of large cardinals (cardinal numbers beyond those for which the existence can be established using only the axioms of ZFC; in this way large cardinals lead to a hierarchy of extensions of ZFC, see \cite{Kan}). It can be shown that the road to superlarge cardinals beyond Kunene inconsitency is open once one accepts more fundamental changes in set theory, contradicting axioms of ZFC (see \cite{DGHJ}). Ulitmately only after reconsideration of all three blocks of axioms jointly (see the previous item) can one decide what is possible. An inconsistency first encountered, can lead to a different approach for the system as a whole and there can still exist e.g. suitable formulations for the foundations of general relativity and quantum mechanics in such an alternate system (The synthetic differential geometry inside certain topoi is also not the same as usual differential geometry in strict axiomatic terms but it can still to a large extent surve the same purpose).     

\end{itemize}

\bigskip

We should observe at this point that even if the more conservative approach of \cite{HH} turns out to be sucessful, we have actually a very similar picture concerning the relation to the foundations of mathematics. In order to work, the approach necessarily requires 

\begin{equation}
P \not = NP
\end{equation}    

to hold. The very same three possibilities as for the case of $P=PSPACE$ arise, here. The only difference is that most experts in the field believe (4) to be derivable from the axioms of ZFC but this only means that the chances for the three outcomes are weighted differently -- if you make a poll amomg experts -- but this is the only difference, otherwise the situation is compeletly the same. \\

\bigskip

Approaches to physics which are based on requirements such as $P=PSPACE$ or $P \not = NP$ (if these requirements can not be derived from ZFC or are even in contradiction to some of the ZFC axioms) mean that the deeper transcendental world of mathematics (beyond those concepts which we use in most parts of mathematics or in the formulation of general relativity, quantum mechanics, gauge theory, etc.) would be shaped by physics based requirements on the hierarchy of complexity classes. \\
This would from a logical perspective \textit{in no way} be different from the way in which counting shaped the most basic axioms of set theory (as we see from the deep role played by Peano arithmetics -- as exemplified e.g. in the techniques behind the G{\"o}del incompleteness results -- or from the cardinal hierarchy). The only difference is of the historical kind: The concept of counting was shaped in ancient Babylonian times and we can no longer truely imagine the transition from using different expressions for \textit{three oranges} and \textit{three stones} to the abstract concept of the number 3. But it is not astonishing at all if modern physics -- e.g. quantum theory -- does influence our further understanding of mathematical concepts. To the contrary, the \textit{unreasonable effectiveness of mathematics} should not -- as Wigner has already stressed -- lead us to expect blindly that our concepts, shaped by experience in the Savannah, should hold without any influence of further experience (as e.g. gained in the quantum world).\\
In the following two sections, we will take a look at further evidence for the relevance of quantum mechanics to the set theoretic foundations of mathematics. 

\bigskip

{\bf Remark:} As is discussed in \cite{Sus}, on very long time scales -- as they take Alice to prepare a firewall at the black hole of Bob -- the wormhole from Alice to Bob -- which allows Allice to send the firewall -- opens up more and more, finally becoming traversable. Now, use the $ER=EPR$ hypothesis. Then causal communication between the partners of the $EPR$ pair should become possible as the wormhole becomes traversable. In nonlocal hidden variable formulations which are equivalent to usual quantum mechanics -- such as Bohmian mechanics -- the consitency of $EPR$ correlations with the causality requirements of special relativity is assured precisely by the nonlocality, which puts the relation between the $EPR$ partners purely on the level of a correlation without any causal communication between them. It is therefore natural to suspect that at the time causal communication between the $EPR$ partners becomes possible, a \textit{local} hidden variable formulation becomes available for the situation. Indeed, this should precisely be provided by the $ER=EPR$ hypothesis. As long as the wormhole is nontraversable this hypothesis states an equivalence between the two horizon conformal field theories at the two ends of the wormhole. But once the wormhole becomes traversable and large enough, there is obviously a completely classical space-time description available for the complete throat as a classical solution of general relativity. We suspect that this classical gravity description is functioning as a \textit{local} hidden variable description for the $EPR$ pair. But a local hidden variable description is -- as is well known from the famous theorem of von Neumann -- not allowed for by quantum mechanics. This should be the way to see the inconsitency of the opening up of the wormhole from the quantum mechanical side -- i.e. the EPR side -- while it is seen by the causal problems traversable wormholes pose by allowing to be used as a time machine from the gravity -- i.e. ER -- side.\\
It is this view on the $ER=EPR$ hypothesis which -- if it could be established by proof -- would lend support to the idea that there might exist a reformulation of the firewall paradox completely in terms of quantum mechanics and classical thermodynamic observers, bringing it close to the setting of Schr{\"o}dinger' Cat and Wigner's Friend. On the other hand, in the setting of \cite{AW} traversable wormholes would precisely provide the closed time-like curves which make the information provided by classical and quantum Turing machines compatible. In this case, the availability of a local hidden variable description -- essentially providing a dual classical description for quantum mechanics -- should no longer cause a problem but it would come along with the prize of a change in set theoretic foundations, e.g. by incorporating a postulate like $P=PSPACE$.\\
For an overview to what extent set theory is not determined by e.g. usual analysis and future axioms might be determined by external influence -- including a discussion on quantum mechanics and hidden variable theories -- see \cite{Mag} (for a detailed example on hidden variables, see \cite{FM}).

\section{Chaitin incompletness}

We have mentioned the difference, in information accessible at any time to a classical versus a quantum system, repeatedly above. Randomness in quantum mechanics exemplifies this point very clearly. In quantum mechanics we encounter \textit{true} randomness (as opposed to the output of a random algorithm) and this seems to be essential for keeping quantum mechanics in line with the causality requirements imposed by the special theory of relativity. It is the true randomness in quantum mechanics which makes sure that for EPR correlated pairs of particles the information in the correlations is the \textit{only} information. If the randomenss in quantum mechanics would e.g. be of the type of the output of a random algorithm there would be additional hidden information -- on the algorithm -- in the individual outcomes which would probably imply superluminal signaling between the partners of the pair. \\
Does the notion of randomness in quantum mechanics agree with the randomness concept used in algorithmic information theory (see e.g. \cite{Cha})? The outcome of a quantum mechanical measurement should be random in the sense that no quantum mechanical system can decode information from it, i.e. no quantum computer should be able to compress the numbers (since otherwise the outcome would not be random in a way compatible with the principles of quantum mechanics itself). Since the notion of general compressability agrees for quantum and classical Turing machines, this means that the outcome should be random in the sense of algorithmic information theory.\\
There is a formulation of G{\"o}del incompleteness in terms of complexity (see \cite{CC} and \cite{Cha}; full equivalence to G{\"o}del incompleteness depends on a small technical assunption on the axiom system $S$ which we neglect, here): Let $S$ denote a formal axiomatic system (e.g. the ZFC system of set theory) and $K(x)$ the Kolmogorov entropy of a string $x$ of symbols in the description language. Then we have the following theorem (Chaitin incompleteness theorem):

\bigskip

There exists a constant $L$ (which only depends on $S$) such that there does not exist a string $x$ for which the statement 
\begin{equation}
K(x) \geq L
\end{equation}
can be proven within the axiomatic system $S$.

\bigskip

This means that for binary numbers whose lenght exceeds $L$ digits, we can not prove their randomness in the axiomatic system $S$. But the randomness postulate of quantum mechanics \textit{requires} randomness for measurement outcomes, related to numbers of arbitrary length. This means that the randomness postulate lives -- as an  example of G{\"o}del incompleteness -- outside the system ZFC. But more so, since any extension of ZFC to a more powerful axiom system $S$ would result in a similar number $L(S)$ -- which would again be surpassed by the randomness postulate -- we have to conclude that the power of the randomness postulate of quantum mechanics is even beyond \textit{any} finitely axiomatisable system $S$. The randomness postulate even claims the randomness of measurement outcomes for \textit{any} quantum mechanical system, i.e. it involves a recourse to any (separable) Hilbert space and therefore a recourse to an infinite family of transfinite structures. From the perspective of axiomatic set theory it is therefore way beyond what is usually used and of a strange circular character (with its back reference to transfinite structures).
In conclusion, the randomness postulate, together with the Chaitin version of G{\"o}del incompleteness, gives another hint that quantum mechanics implies a mathematical framework way beyond the ZFC system.

\bigskip

Let us close this section with a small side remark. Naively, one would expect the requirement of an agreement of classical and quantum computability and imformation, resulting in the requirement $P=PSPACE$, as a stronger requirement than the randomness postulate of quamtum mechanics since it should result in usual quantum theory as a special limiting case. On the other hand, (3) represents only one axiom while -- as we have seen above -- the randomness postulate goes beyond any finitely axiomatizable system. \\
But this is not necessarily a contradiction since the question if we can view a statement as one axiom (or e.g. infinitely many) depends on the language/syntax underlying the formulation of the axiom system. So, this only shows the possible scope of modifications (one has to consider all the blocks of logical and set theoretic axioms as a whole, as we have noted above) imposed by the requirement of (3). \\
\textcolor{blue}{There might be a different perspective possible. As discussed above, it might be possible to see wormholes as providing -- through $ER=EPR$ -- a local hidden variable description. If this would be true, it could mean that in the setting of a theory with $P=PSPACE$ the randomness postulate of quantum mechanics might not be needed any more (since there would exist a dual hidden variable description). This would make a postulate like (3) even more interesting since it seems to be much more managable -- than the randomness postulate -- from a set theoretic perspective (and much closer to postulates set theorists can provide experience with and, hence, give new technical input to the mathematical physics side).}

\section{The model theoretic results of Benioff}

Indeed, a completely rigorous and detailed mathematical analysis of the relationship of quantum mechanics to the foundations of mathematics -- in the form of set theory -- exists (see \cite{Ben1976a}, \cite{Ben1976b}). What Benioff showed is that formulating randomness in quantum mechanics in precise set theoretical terms and requiring a strong enough randomness notion to be satisfied, it is not possible to formulate quantum mechanics in a minimal model of ZFC (or also certain types of extensions of minimal models, as shown in the second paper). One might justifiably claim that these models are very strange (and very restrictive) models, far away from the standard models of ZFC. But the essential point is that the results of Benioff show that quantum mechanics does definitely not work in all models of ZFC. But this is silently assumed if we view ZFC as the foundational basis of mathematics and mathematical physics since we assume in this way that we could in principle at any time reduce all our proofs to manipulations of the axioms of ZFC alone (even if we never do this in practice and have never done it for most of our arguments). If only one model is excluded as a home for quantum mechanics this means that this is not true and that quantum mechanics is living beyond the axiomatic scope of ZFC. \\
In a first step one could be tempted not to view this as a serious problem, taking the standpoint that it just means we have to use some form of extension of ZFC to give a proper formulation of quantum mechanics or, alternatively, accept the intrusion of some pieces of model theory into mathematical physics. But the adaptation of model theory means that silently we assume a meta level to be attached to our theory, the level on which the sets live we build our models from. We say silently because usually it is not necessary to deal with this level with heavy machinery because for most cases a very moderate level of properties of these meta level sets is used (often referred to as naive set theory, for this reason). But wanting to formulate quantum mechanics properly, the situation is very different. At the meta level we talk \textit{about} the axiomatic set theory and \textit{about} quantum mechanics founded in it. But for quantum mechanics this means the meta level is the level of the observer, making use of quantum mechanics and quantum experiments. But -- as in the case of experiments -- suppose we decide to pass to a quantum mechanical description of the observer himself (for the prize of introducing another second level observer). This means that we can no longer treat the meta level with naive set theory because -- as the results of Benioff show -- there will arise situations which are intricate enough not to work in all models of ZFC. In other words, we are not only forced to introduce the full fledged machinery of ZFC but also model theory on the meta level. The latter implies that we need a meta meta level on which to build the models for the meta level from a naive set theory. But now we can iterate the argument, passing to a quantum mechanical description on the meta meta level. In consequence, we are driven into an infinite hierarchy of model theories which once again shows that the randomness postulate of quantum mechanics leads us beyond \textit{any} finitely axiomatizable system.

\bigskip

As a side remark, let us note that the results of \cite{Nab} show that for \textit{any} finitely axiomatizable extension of ZFC -- to be used as foundations for mathematics and mathematical physics -- there remains an indeterminacy of the partition function of Euclidean quantum gravity which is due to G{\"o}del incompleteness. 

\bigskip

Let us close this section by two additional remarks:

\bigskip

\begin{itemize}

\item Dealing with model theory, incompleteness, etc., at the meta level, it is natural to pose the question if there is a relation to the hierarchy of super $\Omega$s of Chaitin. If this could be shown to be the case, it would again show the vastness of modifications in the foundations of mathematics, required by quantum mechanics.

\item Observe that the meta level is precisely the level at which Wigner's Friend lives, leading -- in a conventional classical versus quantum description -- to the incompatibility of recurrence times, above.

\end{itemize}

\section{Conclusion}

Suppose paradoxes in physics (like Schr{\"o}dinger's Cat, Wigner's Friend, or the Firewall Paradox) would force us to adopt an axiom like $P=PSPACE$ (as we have pointed out, the situation is -- in principle -- not different for the $P \not =NP$ case). We collect a few implications this could or would have. 

\bigskip

\begin{itemize}

\item Physics -- once again -- would be at the point of requiring new mathematics, but this time far beyond the invention of calculus by Newton to have the machinery for the formulation of classical mechanics. Similar situations have occured in history. Simple rules of geometry, as e.g. the law of Pythagoras, have been known to Babylonean geometers as heuristic rules but required a thorough foundation of these rules as laws to develop geometry as a mathematical discipline. Indeed, this transition involved the invention of much of the foundational features of mathematics, including the concept of mathematical proof. Another example from history is the concept of number. It is difficult to truely inagine for us today the transition (which took place in Babylonian times) from concrete numbers -- three oranges or three stones -- to the abstract concept of 3. But once this was a change in foundational concepts. The concept of number is still very much behind the way we form the concepts for the transfinite world, as most explicitely seen in the concept of cardinality. Finite sets -- numbers in disguise -- are the prototype we have in mind, when trying to keep some of their properties by invoking appropriate axioms for transfinite sets. This shows very clearly how much our perception of transfinite concepts is shaped by very early experience of human kind in the everyday world. As we have stressed above, there is no reason to expect that our modern eperience with e.g. the quantum world should not influence the way we think about abstract transfinite concepts. Indeed, we invent transfinite concepts as a means to transport our abililty to analytically understand the world way beyond our everyday experience. We should not be surprised if we meet obstacles on this route which form our further perception of the transfinte world by new external input.

\item With postulates like $P=PSPACE$ (or $P \not = NP$) an approach to the foundations of mathematics, influenced by concepts from quantum mechanics, would no longer remain an elsusive subject. Approaches based on quantum logic (trying to introduce a from of quantum set theory) have always suffered from the same problem which we have encountered above for model theory. They need a meta level which either remains classical or leaves us with an unmanageable infinite hierarchy of theories. In contrast, the above postulates are completely in line with the way set theorists approach the transfinite world beyond ZFC (as examples, take Kunen inconsistency or IST -- both mentioned above -- or non-well founded sets). Also, complexity theory -- with the hierarchy of complexity classes -- is a well developed mathematical subject with established foundational contact. 

\item Finally, introducing an axiom like $P=PSPACE$ for the description of the transfinite world, could have profound implications for our understanding of complexity in general. There are processes like protein folding which seem to be far beyond the computational complexity accessible by a quantum computer, i.e. it is completey unclear how nature manages to complete these processes on any reasonable time scale (see \cite{DD}). A list of such examples could be continued, including e.g. topics from material science. Mathematical foundations based e.g. on (3) could therefore lead to the development of the appropriate calculational tools for these processes. In this way, they could even have huge economical impact. The availability of the calculational tools of calculus was of tremendous importance for the development of steam engines and the subsequent quest for higher efficiency -- leading finally to the internal combustion engine -- the development of railway lines, etc. in the 18th and 19th century. The basic concept of a steam engine was already known in Greek antiquity but it was probably very much the availability of the appropriate calculational tools which decided between having an amusimg spectacle for puppet theatres or an industrial revolution.     

\end{itemize}  

\appendix

\section{A physics perspective}

The Firewall Paradox has the character of a \textit{causal} paradox. The decoding of the Hawking radiation by Alice makes Information available to infalling Bob which should not yet be available to him. To a considerable extent, general relativity has the character of a thermodynamic theory, with e.g. concepts like entropy, enthalpy, and temperature all playing their role in black hole physics. In the example of Schr{\"o}dinger's Cat (and the paradox of Wigner's Friend, building on it) we have -- in the view presented above -- a \textit{purely} thermodynamic decoding problem giving rise to the paradox. Could it be that the Firewall paradox helps to translate -- by making use of the fact that space-time in general relativity plays a double role with many concepts allowing for a thermodynamic, as well, as a causal space-time view -- a clash between quantum mechanics and a classical thermodynamic world of observers -- as exemplified by Schr{\"o}dinger' Cat and Wigner's Friend -- into a causal paradox? We have discussed above that this clash between quantum mechanics and classical thermodynamic observers appears largely to be rooted in a difference of information available at a given time. In the firewall paradox it is also information which should classically not yet be available to Bob which causes the paradox. \\
Normally, the thermodynamic character of a theory means that it is not quantized at all but is taken to signal the existence of a microscopic theory behind it and it is only the microscopic theory which gives proper rise to the quantum description. In the case of general relativity, this is precisely the perspective which string theory provides. What is not clear is if such a perspective also works beyond phenomena which can be approached with local experiments. In general relativity, the Planck scale provides a limt of what can be investigated with local experiments. But as shown in \cite{Sus}, the far ultra-Planckian regime and length scales far below the Planck length might well have physical meaning, albeit not in the sense of information provided by local experiments but through observation on very long time scales. In the view presented in \cite{Sus}, the Planck scale is merely a complexity bound and the very high degrees of complexity of the ultra-Planckian regime can only be investigated on very long time scales (since they are in non polynomial complexity classes with a quantum computation taking a very long time).\\
For these high complexity phenomena beyond the Planck scale it is not a priori clear if we can conclude the existence of a microscopic theory -- giving in turn rise to a proper quantized description -- from the thermodynamic character of general relativity. 

\bigskip

Is it conceivable that string theory could play the role of a realization of a computability requirement for degrees of freedom below the Planck scale (and, hence, in the polynomial complexity class)? Following the view on the Planck scale as a complexity bound, the degrees of freedom observed below this bound are precisely those which can be accessed by a quantum computer in short -- i.e. polynomial -- time. This is the reason why they are accessible by (space-time) local -- in contrast to long time -- experiments. With length scales far below the Planck length (i.e. degrees of freedom far above the Planck scale, understood as an energy scale) still having a physical meaning in general relativity (see \cite{Sus}), this means that general relativity should inherently endow the degrees of freedom at length scales above the Planck length with a structure which makes them accessible to quantum computation. Could string theory be viewed as providing a realization of this quantum computability requirement? \\
In this view, the quantum computability requirement would be the abstract concept and string theory (one) concrete realization, i.e. the relationship of an abstract complexity hierarchy/quantum computing based approach to string theory would be roughly reminding of the relationship of abstract Hilbert space quantum mechanics to the concrete realization in wave mechanics. It is interesting to note in this respect that the Leech lattice -- which is used to obtain the compactification torus for bosonic string theory, leading to the monster vertex algebra -- has a deep connection to the Golay error correcting code (which was used e.g. by NASA in the Voyager missions).

\bigskip

Viewing ultra-Planckian degrees of freedom in general relativity as non-accessible by quantum computation (in polynomial time) would mean that for these degrees of freedom we can not resolve the thermodynamic character of the theory by an introduction of micro states which then give rise to a proper quantized theory. The quantum micro states would simply be non accessible by quantum computation, i.e. non accessible as proper quantum mechanical systems. But if the thermodynamic character of general relativity can not be resolved in this way for all degrees of freedom, the question of making general relativity compatible with quantum mechanics acquires a new degree of urgency because it is normally not possible to quantize a thermodynamic theory at face value. This -- and the possible physical meaning of length scales in general relativity far below the Planck length in long time observations -- brings us back to the old question if general relativity should be quantized at all. \\
The most convincing argument that it definitely has to be dates back to Feynman, using the following argument: Suppose we consider the usual double slit experiment for electrons. The usual argument why we can not simply check which of the two slits the electron passed through while retaining the interference pattern at the same time, relies on the quantization of electromagnetic radiation. Using a photon send to the double slit to measure the position of the electron, we have to use a sufficiently short wave length to allow -- by the usual rules of optics -- for a resolution below the dimensions of the double slit. But since the elctromagnetic radiation is quantized we can not decrease the intensity below the one photon level, leaving us with a minimum of energy and momentum of the radiation. The subsequent momentum transfer to the electron in the course of measurement leads to the well known uncertainty in the electron momentum, consequently destroying the interference pattern. If gravity would not be quantized, we could simply use gravitational waves instead of electromagnetic radiation for the position measurement of the electron at the double slit. Since the gravitational waves would not be quantized, we would be free to decrease the intensity of the beam arbitrarily. Hence, we would not disturb the momentum of the electron leaving us with an interference pattern while knowing the position of the electron -- to arbitrary accuracy -- at the same time. The important point is that this double slit experiment is \textit{not} some quantum gravitational version of a double slit. It is a usual double slit experiment with electrons. The conclusion is that a non quantized theory of gravity would destroy the consitency of quantum mechanics even in those cases in which it is very well experimentally established. \\
But the argument of Feynman applies in an obvious way only to degrees of freedon which can be accessed by local experiments (like measuring a precise position at a \textit{given} time). The compatibility of these degrees of freedom with quantum mechanics would be assured if general relativity satisfies a quantum computabilty requirement for these degrees, as suggested above. This leaves us with only one a priori argument that general relativity as a whole -- including the ultra-Planckian regime -- has to be quantized: The incompatibility of the information content at a given time between classical and quantum theories is deeply involved in all of the paradoxes discussed above, i.e. Schr{\"o}dinger's Cat, Wigner's Friend, and the Firewall Paradox. But this incompatibility is resolved in a setting as in \cite{AW} (giving rice to $P=PSPACE$) with general relativity and quantum mechanics providing the same information content -- in the sense of the same computational accessability -- at any time. In such a setting there appears to be no obvious argument in favour of a quantization of general relativity. \\

\bigskip

Returning to a more mathematical viewpoint, in a foundational change in mathematics, including (3), the issue of the conflict between general relativity and quantum mechanics would be resolved by placing both theories into a foundational set theory framework which ensures the consitency of their information content. Thus general relativity would not need to be quantized in the conventional sense. Placing it into a different foundational framework would be akin -- though different in important respects -- to the way synthetic geometry is introduced as differential geometry inside a topos. \\
Chris Isham has for several decades stressed the fact that it is not obvious at which level of the mathematical concept of manifold structure the transit to making general relativity compatible with quantum mechanics has to take place. Canonical approaches take the metric as the relevant structure, leaving the differential and topological structure of the manifold fixed. More general approaches try to sum also over topologies and (exotic) differential structures. Still, even these approaches leave the underlying set theoretic structure as given. As Isham has stressed, the underlying point set structure has the well known physical meaning of events. In special and general relativity we normally do not think of events as an a priori given structure but as defined only by physical processes (e.g. as the intersection of two world-lines). Hence, it is not obvious that it might not be the level of set theory which requires modification in order to achieve consitency of general relativity with quantum mechanics. What we have suggested above is that the Firewall Paradox might point very concretely into this direction. 

\bigskip

Finally, let us collect a few additional reamarks: 

\bigskip

\begin{itemize}

\item One might pose the question why compatibility of the information -- available at any time --  of classical and quantum systems, as shown for the setting of \cite{AW}, should necessarily lead to a change in the set theoretic foundations of mathematics and mathematical physics. After all, the analysis of \cite{AW} is completely done in the conventional framework of ZFC and the compatibilty results completely from the physical phenomenon of closed timelike curves. In spite of this, the analysis in \cite{AW} does imply that classical \textit{and} quantum Turing machines in this setting can handle problems in the complexity class $PSPACE$ in polynomial time. As for passing from classical to quantum Turing machines, we have a different concept of (classical \textit{and} quantum) Turing machine in the setting of \cite{AW}. For these Turing machines problems in the complexity class $PSPACE$ are polynomial, i.e. viewed \textit{internally} from a set theory -- or indeed any other form of axiomatic theory -- which is compatible with these Turing machines (i.e. proofs in the axiom system of this set theory can be seen as programs running on such a Turing machine, just the same way this is true for proofs in ZFC and conventional Turing machines) the postulate $P=PSPACE$ has to hold. We do not think of such foundational changes of set theory on passing from (usual) classical to (usual) quantum Turing machines for the simple reason that in this case the speed up achieved by quantum computers for certain problems does not seem to seriuosly interfer with the hierachy of complexity classes. In contrast, (3) is by most experts in the field believed to be contradicted by ZFC. We can heuristically use an approach in the conventional setting of ZFC -- as in \cite{AW} -- as a fist step but ultimately the Turing machines which the theory describes and those which the experimental scientist (or the mathematician who constructs the theory, like the external observer an experiment) uses, should agree. Even if one does not believe in a full fledged requirement of the Church-Turing thesis, one can not accept the two concepts of Turing machines to result in contradicting axioms. Turing machines are not different in this respect from clocks, as used in the famous thought experiments founding special and general relativity by Einstein. The Turing machines which the theory internally describes give the predictions of what the theory supposes an observer to measure on a Turing machine and in checking these predictions experimentally on a Turing machine we have to suppose that the two concepts fundamentally agree and are definitely not in mutual contradiction to each other.     

\item While the passage from (usual) classical to (usual) quantum computers does not seem to seriuosly affect the complexity hierarchy (not to the degree leading to obvious candidates for contradiction to the ZFC axiom system), it leaves the concept of proofabilty completely unchanged. In other words, while there is a speed up for certain problems on quantum computers, the halting problem is not affected at all. A program running on a (usual) quantum computer does halt if and only if it does on a (usual) classical Turing machine and if we can prove this on one of the machines we can always on the other. This is not true even in passing to \textit{classical} general relativity (without anything like Hawking radiation involved). As is well known, the class of Malament-Hogarth space-times (to which e.g. the internal part of the Kerr solution belongs) does not respect the halting problem, i.e. an observer in these space-times may be able to observe if a Turing machine does halt or not, even if this is impossible to decide in ZFC (e.g. he might be able to decide about the consistency of the axiom system ZFC itself, in this way; see \cite{Ear} for a general account on Malament-Hogarth space-times and \cite{EM} for the first detailed and conclusive study of the Kerr solution).\\
The scenario of \cite{AW} -- with polynomial computing time for the complexity class $PSPACE$ (for both, classical and quantum computers) -- puts space and time resources in computing on an equal footing. This alone is a sign of a certain naturality of such Turing machines in a relativistic setting. As we have mentioned above, the central ingredient in \cite{AW}, leading to the compatibility of classical and quantum computing power, is the presence of closed time-like curves. The polynomial time accessability of the complexity class $PSPACE$ is a consequence. A priori it is therefore not clear if the postulate $P=PSPACE$ for a foundational change in set theory is sufficient (If not, one would have to search for a more general change of axioms, including or implying $P=PSPACE$. Since this would not change the general character of the approach discussed above, we have decided to restrict completely to the simplest possibility that this postulate alone might be sufficient. In any case, it would provide a natural starting point to discuss foundational changes.). The naturalness of an equivalence of space and time resources in computing in a relativistic setting is a concrete argument in favour of precisely the postulate (3). \\
If a change in foundations, based on (3), is sufficient to include also the increased power with respect to proofability, which Turing machines in Malament-Hogarth space-times show, is an open question for research. Presently it seems that Malament-Hogarth space-times naturally fit into the scenario of \cite{AW} since known examples show the appearance of closed time-like curves -- as is e.g. well known from the interior of the Kerr solution -- and this seems to be a generic feature but no general theorem, establishing this, is known so far (see \cite{Ear}).\\
Closed time-like curves are always viewed with some suspicion in physics, even if a consitency condition is used. But even in the most general case -- without a consistency condition invoked -- we must confess that -- once we take into account the possibility of a change on the foundational set theory level -- it is not clear if their paradoxical character is inherent or arises from using inappropriate foundations. To avoid misunderstanding: We do not imply here that any form of foundational change would not have to respect consitency but what we think of is that placing the geonetry of general relativity into a different foundational framework could avoid the appearance of paradoxes. To give a concrete example: Different outcomes for 2$\pi$ and $4\pi$ rotations are paradoxical from the viewpoint of classical geometry. But they are not for spin-$\frac{1}{2}$ systems in quantum mechanics, simply because geometrically we pass to the group $SU(2)$, thereby killing the first nontrivial homotopy group of the rotation group $SO(3)$, in this way resolving the paradox. Note that interestingly there is a connection to gravity if one pursues this approach furhter in higher dimensions: Killing the next higher nontrivial homotopy group of $SO(n)$ leads one to the string group which is behind Green-Schwarz anomaly cancellation and the story goes on in the next steps, again, with fivebrane anomaly cancellation (relating to the dominating degree of freedom in $M$-theory, if one counts the number of central charges in the 11-dimensional supersymmetry algebra) and ninebrane anomaly cancellation (relating to boundary conditions, respectively domain walls, in $M$-theory and space-filling branes in string theory) appearing(see \cite{SSS} and \cite{Sat}). What we suggest, here, is that something similar might happen not only for classical homotopy groups, i.e. closed loops in classical geomtery, but also relativistically for closed time-like curves.         

\item \textcolor{blue}{The relation of a foundational approach to \cite{HH}:}  Let us close with a -- we think important -- remark. As we have stressed above, the situation in the approach of \cite{HH} is not really different from one using e.g. (3) for a change in the underlying set theoretic foundations. The approach of \cite{HH} requires $P \not = NP$, in order work. While one would certainly weigh the odds differently -- as compared to a postulate like (3) -- for this to be simply derivable in ZFC, at present we simply do not know this. In other words, basing an approach in mathematical physics on \cite{HH}, we have to be prepared to accept that it \textit{could} force us into accepting a change in foundations (and for the very same reasons we have expelled for the case of (3), there is no reason why we should not see this as a natural development of our view on the world of transfinite concepts). But beyond this, though an approach based on (3) appears as an alternative path to \cite{HH}, we think that to a certain extend it relies on the approach of \cite{HH} to work in a first step. The situation is very much the same as we have discussed it above for \cite{AW}. We presently arrive on any of our conclusions about the firewall paradox based on an approach which is founded in ZFC. This only can work if it is not outright inconsistent from the start. This should mean that for situations were we can for the first steps start working with the usual classical and quantum Turing machine concepts -- as opposed to the more involved machines of \cite{AW} -- there should be something like the complexity hierarchy based scenario of \cite{HH} shielding us from outright inconsistency from the start. \\
The situation might be even more extreme if the firewall paradox could be given a formulation purely in terms of quantum mechanics and classical thermodynamic observers. The way it stands, the firewall paradox establishes a clash between quantum mechanics and general relativity, i..e it shows that we have two theories for which it is inconsistent to apply them jointly to the same class of phenomena, even in weak gravitational fields. Suppose, one could give a purely quantum mechanical formulation. Suppose, as the most extreme case, this could be done along the lines of \cite{Ben1976a} and \cite{Ben1976b}, bringing back the problem of accessible information -- lying at the root of the firewall paradox -- to accessible information under the quantum randomness postulate. Hilbert space quantum mechanics alone can be seen to a large degree as a purely axiomatic mathematical discipline. So, a paradox in this setting would come very close to an inconsitency of ZFC. But we know from experience that starting mathematics and mathematical physics from ZFC is -- at the very least -- an extremely successful first step. We could not derive anything sensible about quantum mechanics -- and possible implications it might have for foundational isssues -- if this would not hold for a very large degree. So, if this would happen (to just take the most extreme example for illustration) there would still have to be some mechanism shielding ZFC from outright inconsistency. The use of the complexity hierarchy in the approach of \cite{HH} and \cite{Sus} might provide such a mechanism. In the most extreme case, ZFC could be effectively consistent in the way David Ruelle once coined this term. A proof of inconsistency from within the axiom system ZFC could just take too long -- in its most compressed form in terms of complexity theory -- to be established in cosmological time scales experienced so far in the universe (in this case, our experience from the Savannah would carrry us a long way but not into a Platonic transfinite world). So, for problems of low enough complexity it would be justified to deal with the usual concept of Turing machine but beyond a certain complexity scale we would be forced into accepting a different concept of Turing machine and, hence, a different foundational framework. \\
Even if we do not consider the case of mathematical inconsitency, a mechanism resembling the effective consitency of Ruelle might be relevant. It could just take a very long time -- in the sense of computational complexity -- to establish $P \not = NP$ in $ZFC$ even if it would hold, there. The same could be true for disproving $P=PSPACE$. In this way, using $ZFC$ as a staring point of our considerations might be compatible even if we would ultimately need a different foundational framework. We always need a form of upward compatibilty in physics if we pass to a new and more general framework (e.g. in passing from classical mechanics to special and general relativity or to quantum mechanics) and this would also have to hold true if we would have to change foundations. The approach of \cite{HH} and \cite{Sus} could ensure this if it would be applicable to this effective consitency like scenario. This would mean that it would not only be applicable to computing times but also to proof length for some special questions (like e.g. $P \not = NP$), i.e. for these questions the complexity considerations of this approach would also apply to complexity of proofs itself.              

\end{itemize}

\section{A final mathematical remark}

As we have mentioned above, ultimately Turing machines used internally in the description of a theory (like those in \cite{AW}) and Turing machines used by an observer (or the mathematician formulating the theory) have to agree because the internal concept tells us what the theory expects the observer to measure on his Turing machine (just as in the discussion of clocks in relativity). This even holds true if internal Turing machines differ in terms of proofability (as in the case of Malament-Hogarth space-times). Everything else can only be a first approximation. We have discussed above how it might be possible that such a first approximation works at all. This was based on a kind of effective compatibility of the two foundational theories, much of the type of the effective consistency suggested by Ruelle. Such effective properties of theories are close to the way the theoretical physicist thinks about the relation of his theories to each other (e.g. the concept of effective quantum field theory).\\
There might be a more mathematical way to think about this situation: Discussing e.g. Turing machines \textit{in} general relativity (as in \cite{AW} or in the case of Malament-Hogarth space-times) silently assumes that we can apply general relativity also on the meta level on which -- as we have discussed above in the case of model theory -- normally naive set theory resides. One might argue that the meta level is not truely involved since set theory allows to transform this situation back to a description purely within set theory. Though such an argument might be applied if we discuss only speed ups of Turing machines (as e.g. in the case of quantum computers), it does not seem to be applicable if the internal Turing machines differ in terms of proofability from what we can achive in ZFC (because then -- by definition -- we could never reproduce their proofs in ZFC). In this case, we truely assume the applicability of general relativity on the meta level.\\
Mathematically, a sufficient condition to achieve this is the full applicability of ZFC on the meta level (instead of naive set theory). From a set theoretic viewpoint this gives approaches to Turing machines in Malament-Hogarth space-times the flavor of forcing. It is no contradiction that the internal theories might differ from ZFC. This means that we can not reproduce this situation in set forcing since in this case the new model, resulting from forcing, will again satisfy the ZFC axioms. \textit{If} a technique similar to forcing is applicable to these cases, one would -- at least -- need class forcing where the new model need no longer be one of ZFC. If forcing would be applicable, it would mean that the needs of physicists in approaches like Malament-Hogarth space-times or \cite{AW} could be satisfied within a highly developed framework -- even if more exotic forcing techniques beyond class forcing would turn out to be needed -- and that set theorists would be able to offer the technical expertise for a new tool in mathematical physics. \\
How would the effective theory viewpoint fit into a picture based on forcing (assuming that the latter would be applicable, here)? For the sake of easy illustration let us restrict to the most radical case, here, the effective consistency of Ruelle. Normally, one would take the view that forcing relies on assuming the consitency of ZFC since one constructs the exotic models from a model of ZFC. But consider the following speculative situation: Suppose we would establish inconsistency of ZFC. We would surely go on to isolate those ZFC axioms responsible for the contradiction. Suppose this is axiom $A$. Suppose additionally -- not logically important in the sequel -- that we would have reason (e.g. from physics) to view the contradiction as \textit{natural}. Suppose further that we would have constructed by forcing a model contradicting $A$ (using $\neg A$, as most experts would suspect to be the case for a model with $P=PSPACE$). Since the model anyway uses $\neg A$, intuitively we would have \textit{no} reason to suspect that it is also inconsistent (in spite of being constructed from a ZFC model). This would be even more convincing -- though, again, logically not necessary -- if we would have reason from physics (as e.g. in the case of $P=PSPACE$) to construct this model for a more advanced theory, i.e. if we would view this model from a physics based perspective as the superior one, anyway. \\
Forcing (class forcing and more general forcing notions) should in this way naturally include a \textit{mathematical version} of effective consistency (as formulated by Ruelle on a physics/cosmology basis). If e.g. the forcing construction of the new model itself -- not involved arguments within it -- does not exceed proof length $L$ and we could show that a proof of inconsistency of ZFC needs proof length $\gg L$, we would have a version of effective consistency.

\bigskip

\textcolor{blue}{Finally, we should stress that much of our discussion on a possible interaction of physics to foundational questions in set theory, here, is not limited to the special case of theories with $P=PSPACE$ (we have repeatedly mentioned above that the situation with an approach as followed in \cite{HH} and \cite{Sus} is -- concerning the necessary $P \not =NP$ requirement -- not really different, \textit{in principle}) and even not to the firewall paradox. When discussing Turing machines in Malament-Hogarth space-times -- a setting which is from the physics side placed completely in the well established theory of general relativity -- physicists have silently assumed the applicability of general relativity on the meta level, i.e. they have defintitely left the framework of ZFC proper. It is in any case an interesting question -- for both physics and set theory -- how to put this onto a mathematical basis which is rigorous also from the high standards of axiomatic set theory. Questions, like the applicability of a framework like (class or more exotic) forcing for this, arise naturally, here.}

\end{document}